\documentclass[a4paper,12pt]{article}
\usepackage{epsfig}
\newcommand{\bmat}{\left(\begin{array}}
\newcommand{\emat}{\end{array}\right)}

\def\lsim{\raise0.3ex\hbox{$\;<$\kern-0.75em\raise-1.1ex\hbox{$\sim\;$}}}
\def\gsim{\raise0.3ex\hbox{$\;>$\kern-0.75em\raise-1.1ex\hbox{$\sim\;$}}}

\def\yzero{\smash{\hbox{$y\kern-4pt\raise1pt\hbox{${}^\circ$}$}}}

\def\s2{\frac{1}{\sqrt2}}

\def\beq{\begin{equation}}
\def\eeq{\end{equation}}
\def\beqa{\begin{eqnarray}}
\def\eeqa{\end{eqnarray}}

\def\IF{\relax{\rm I\kern-.18em F}}
\def\II{\relax{\rm I\kern-.18em I}}
\def\IP{\relax{\rm I\kern-.18em P}}
\def\IC{\relax\hbox{\kern.25em$\inbar\kern-.3em{\rm C}$}}
\def\IR{\relax{\rm I\kern-.18em R}}

\def\Dsl{\,\raise.15ex\hbox{/}\mkern-13.5mu D} 
\def\IZ{Z\kern-.4em  Z}
\def\bmat{\left(\begin{array}}
\def\emat{\end{array}\right)}

\def    \part          {\partial}
\def    \be            {\begin{equation}}
\def    \ee            {\end{equation}}
\def    \bea           {\begin{eqnarray}}
\def    \eea           {\end{eqnarray}}

\begin{document}
%
\pagestyle{empty}
\rightline{CERN-TH/2001-299}
\rightline{FTUAM 01/20}
\rightline{IFT-UAM/CSIC-01-33}
\rightline{hep-ph/0110381}
\rightline{October 2001}

\renewcommand{\thefootnote}{\fnsymbol{footnote}}
\setcounter{footnote}{0}

\vspace{1.5cm}
\begin{center}
\large{\bf A kind of prediction from superstring model building
\\[5mm]}
\vspace{1cm}
\mbox{\sc{
\small{
C. Mu\~noz
}
}
}
\vspace{0.5cm}
\begin{center}
{\small
{\it 
Theory Division, CERN,
CH-1211 Geneva 23, Switzerland. \\
\vspace*{2mm}
\it 
Departamento de F\'{\i}sica
Te\'orica C-XI, Universidad Aut\'onoma de Madrid,\\[-0.1cm]
Cantoblanco, 28049 Madrid, Spain. \\
\vspace*{2mm}
\it 
Instituto de F\'{\i}sica Te\'orica  C-XVI,
Universidad Aut\'onoma de Madrid,\\[-0.1cm]
Cantoblanco, 28049 Madrid, Spain.\\
} 
}
\end{center}

\vspace{1.5cm}

{\bf Abstract} 
\\[7mm]
\end{center}
\begin{center}
\begin{minipage}[h]{14.0cm}
Assuming that the Standard Model of particle physics arises
from the $E_8\times E_8$ Heterotic String Theory,
we try to solve the discrepancy between the
unification scale predicted by this theory 
($\approx g_{GUT}\times 5.27\cdot 10^{17}$ GeV) 
and the value deduced from LEP experiments 
($\approx 2\cdot 10^{16}$ GeV). A crucial ingredient in our solution
is the presence at low energies of 
three generations of supersymmetric Higgses
and vector-like colour triplets. As a by-product our analysis
gives rise to a strategy which might be useful 
when constructing realistic models.

\end{minipage}
\end{center}
\vspace{0.5cm}
\begin{center}
\begin{minipage}[h]{14.0cm}

\end{minipage}
\end{center}
\newpage
\setcounter{page}{1}
\pagestyle{plain}
\renewcommand{\thefootnote}{\arabic{footnote}}
\setcounter{footnote}{0}
%
%

\section{Introduction}

The Standard Model of particle physics \cite{Glashow} provides
a correct description of the observable world. However, 
there exist
strong indications that it is just a low-energy effective theory.
There is no answer in the context of the Standard Model to some
fundamental questions.
For example, How can we unify it with
gravity? And then: How can
we protect 
the masses of the scalar particles against quadratic divergences
in perturbation theory (the so-called hierarchy problem)?
Other questions cannot even be
posed:
Why is the Standard Model gauge group
$SU(3)\times SU(2)\times U(1)_Y$?
Why are there three families of particles?
Why is the pattern of quark and lepton masses so weird?

Supersymmetry \cite{susy} is an interesting step in answering
some of these questions. In addition to introducing a kind of unification
between particles of different spin, it also contributes to solving
the hierarchy problem of the Standard Model. It ensures the stability
of the hierarchy between the weak and the Planck scale.
Furthermore, its local version, Supergravity \cite{sugra}, leads to a partial
unification of the Standard Model with gravity.
However, only String Theory 
has the potential to unify all gauge interactions 
with gravity \cite{Scherk} in a consistent 
way \cite{Green}. In fact Supergravity is the low-energy limit
of (Super)string Theory. 
In this sense Superstring Theory would be the
fundamental theory of particle physics which,
in principle, might be able to answer all the above questions.
Its detailed analysis is therefore very important.

One crucial step in this analysis
consists in achieving contact between the theory and the low-energy
world. We need to find a consistent Superstring Theory in four dimensions
which is able to accommodate the observed Standard Model of particle
physics, i.e. we need to find the Superstring Standard Model.
In the late eighties,
the compactification of the $E_8\times E_8$
Heterotic String \cite{Gross} on six-dimensional orbifolds \cite{Dixon}
proved to be an
interesting method to carry out these 
tasks \cite{Wilson}--\cite{grand}
(other attempts at model building, using Calabi--Yau spaces
\cite{Candelas} and fermionic constructions \cite{Bachas}, 
can be found in refs. \cite{Ross} and \cite{Ellis}--\cite{Lykken}, 
respectively).

It was first shown that the use of background fields 
(Wilson lines) \cite{Dixon,Wilson} on the torus defining the
symmetric $Z_3$ orbifold can give rise to four-dimensional
supersymmetric models with gauge group
$SU(3)\times SU(2)\times U(1)^5\times G_{hidden}$
and three generations of chiral particles with the correct
$SU(3)\times SU(2)$ representations (plus many extra particles)
\cite{Kim,Mas}. 
In fact, it was also shown that the three generations
appear in a natural way using just two discrete Wilson lines.
This is so because in addition to the overall factor of 3 coming from
the right-moving part of the untwisted matter, the twisted matter
come in 9 sets with 3 equivalent sectors on each one, since there
are 27 fixed points. In this way, all matter (including extra
particles) in these constructions appear automatically with three
generations.

The next step was the calculation of the $U(1)$ charges
and the study of the mechanism for anomaly cancellation
in these models \cite{Katehou}, since an 
anomalous $U(1)$ is usually present after 
compactification \cite{FayetIliopoulos}.
This allowed the construction of combinations of the non-anomalous $U(1)$'s
giving the physical hypercharge for the particles of the Standard
Model,
although it was also found that the hidden sector is, in general,
mixed with the observable one through the extra $U(1)$ charges.
Fortunately, it was also noted that the 
Fayet--Iliopoulos D-term \cite{FayetIliopoulos}, 
which appears because of the presence of 
the anomalous $U(1)$, can give
rise to the breaking of the extra $U(1)$'s and,
as a consequence, to the hiding of the previously mixed
hidden sector \cite{Katehou,degenerate,Casas1}.
This is because, in order to preserve supersymmetry at
high energies, some scalars with $U(1)$'s quantum numbers 
acquire large vacuum expectation values (VEVs).
It is worth noticing here that 
the Fayet--Iliopoulos D-term was also proposed in order to produce
inflation in these models \cite{inflation}.

In this way it was possible to construct supersymmetric 
models (or, more precisely,
vacuum states) where the original 
$SU(3)\times SU(2)\times U(1)^5\times SO(10)\times U(1)^3$ 
gauge group \cite{Kim}
was broken to
$SU(3)\times SU(2)\times U(1)_Y\times SO(10)_{hidden}$ 
with three generations of particles in the observable sector with the correct 
quantum numbers \cite{Casas2,Font,Katehou}. In addition, baryon- and 
lepton-number-violating operators are absent.
Unfortunately, we cannot claim that one of these models is the 
Superstring Standard
Model. It is true that, as we will discuss below, 
the initially large number of extra particles is 
highly reduced 
since many of them get a high mass 
($\approx 10^{16-17}$ GeV) through the
Fayet--Iliopoulos mechanism, 
thus disappearing from the low-energy theory
\cite{Casas2,Font,Katehou}.
However, in general, some extra $SU(3)$ triplets, $SU(2)$ doublets and 
$SU(3)\times SU(2)$ singlets still remain.
On the other hand, given the predicted value for the unification scale in
the Heterotic String \cite{Kaplu}, 
$M_{GUT}\approx g_{GUT}\times 5.27\cdot 10^{17}$ GeV,
with $g_{GUT}$ the unified gauge coupling,
the values of the gauge couplings deduced from
CERN $e^+ e^-$ collider LEP experiments cannot be obtained \cite{rges}.
Finally, it was not possible to obtain
correct Yukawa couplings in these
models \cite{Katehou,Font,Casas3}.

In any case, it is plausible to think that another orbifold model
could be found with the right properties. In the present paper
we will adopt this point of view. 
We will assume that the Supersymmetric Standard Model 
arises from the Heterotic String compactified on a 
$Z_3$ symmetric orbifold with two Wilson lines.
Then, we will try to deduce the phenomenological properties 
that such a model must have
in order to solve the first and second problems mentioned above,
namely extra matter and gauge coupling unification\footnote{As a 
matter of fact,
our arguments will be general and can be applied to any
scheme giving rise to three generations, since
extra matter and anomalous $U(1)$'s are generically present
in compactifications of the Heterotic String.}.
In fact, the two problems are closely related, since
the evolution of the gauge couplings from high to low energy
through the renormalization group equations 
depends on the existing matter \cite{mass,Giedt2}. With our solution 
we will be able to predict
the existence of three generations of supersymmetric Higgses
and vector-like colour triplets
at low energies. As a by-product, a strategy 
to construct the sought-after Superstring Standard Model
will arise.

Concerning the third problem, how to obtain
the observed structure of
fermion masses and quark mixing angles,
this is
the most difficult task in string model building,
and beyond the scope of this paper.
However, we will mention it briefly in the conclusions.
Needless to say, the experimental confirmation of neutrino masses in the
near future will make this task even more involved.

Finally, before starting with our computation,
let us mention that in the late nineties it has been
discovered that explicit models, with interesting phenomenological properties, 
can also be constructed using D-brane configurations from 
type I string vacua \cite{dbranes}. 
It might well be possible that the Superstring Standard Model arose
from one of them.
However, in our opinion, those models
are still not as satisfactory as the (already fourteen years old)
heterotic ones described above.


\section{Scales of the theory}

Since we are interested in the analysis of gauge couplings,
we need to first clarify which are the relevant scales for
the running between the mass of the $Z$, $M_Z$, and the unification point.
Two of them are quite clear in heterotic compactifications.
We have first the different thresholds associated to the
masses of the supersymmetric particles. 
For the sake of simplicity we will assume that supersymmetry remains
unbroken for energies above 500 GeV, and we will use this value
as our overall supersymmetric scale $M_S$.
In fact, we will use a similar approximation
for the thresholds due to the top quark and light Higgs doublet,
since we will take $M_Z$ as our overall non-supersymmetric scale.
The second relevant scale is associated to the Fayet--Iliopoulos
mechanism to be discussed now. 

As mentioned in the Introduction, some scalars,
in particular $SU(3)\times SU(2)$ singlets, develop VEVs in order
to cancel the Fayet--Iliopoulos D-term. 
This is given by \cite{FayetIliopoulos}:
%
\bea
D^{(a)}=\sum_{i} Q_i^{(a)} \eta_i \frac{\partial K}{\partial \eta_i}
+ \frac{g_{GUT}^2\ tr Q^{(a)}}{192\pi^2} {M_{P}^2}
\ ,
\label{FI}
\eea
where 
$\eta_i$ are the scalar fields with charges $Q_i^{(a)}$
under the anomalous $U(1)$, $K$ is the K\"ahler potential 
(e.g. considering an overall modulus 
$T$, $K=(T+\overline T)^{-n} \eta_i \eta_i^*$ with $n=1,2,3$ for
untwisted, twisted non-oscillator and twisted oscillator fields, respectively),
and $M_P=M_{Planck}/\sqrt{8\pi}$ is the reduced Planck mass. 
Obviously tr $Q^{(a)}=0$ if the model does not have any anomalous 
$U(1)$, a situation that is not very common in $Z_3$ orbifold 
constructions, as we will see below.
Since we want to have a 
vacuum state preserving the physical hypercharge $Y$, we
have to look for the subset of singlet fields with vanishing 
$Y$-charges. 
Remarkably enough, in all
constructed models there is a large subset of singlets,
say $\chi_j$, with 
$Y=0$. 
Although their VEVs are model dependent, as we can see
from eq. (\ref{FI}), an estimate can be done
with the average 
result $\langle\chi_j \rangle\sim 10^{16-17}$ GeV 
(see e.g. ref. \cite{Giedt2}).
After the breaking, many particles, say $\xi$, acquire a high mass because
of the generation of effective mass terms. These come for example
from operators of the type $\chi_j\xi\xi$.
In this way vector-like triplets and doublets and also singlets become 
very heavy. 
We will see
that this is the type of extra matter that typically 
appears in orbifold constructions. Again, for the 
sake of simplicity, we will use the above value as our overall
Fayet--Iliopoulos scale $M_{FI}\approx 10^{16-17}$ GeV.

In principle, other thresholds might appear in these
Heterotic String constructions.
These would be due to the possible presence of 
higher order operators. For example, we might have terms in the
superpotential
of the type $\frac{1}{M_P^{m-1}}\chi_1\cdot\cdot\cdot\chi_m\xi\xi$, which would
produce masses of the order of $(M_{FI}/M_P)^{m-1} M_{FI}$.
Therefore, depending on $m$, intermediate scale masses might be
generated.  
Obviously, the presence of particles with these masses
is very model-dependent and introduces a high degree of uncertainty
in the computation. However, it is important to remark that
the presence of the above non-renormalizable
couplings is not always allowed in string constructions. 
First of all, they must be gauge-invariant, something that is not
easy, because of 
the large number of $U(1)$ charges associated to the
particles in these models.
Even if the couplings fulfil this condition, this does not mean that
they are automatically allowed. They must still fulfil   
the so-called `stringy' selection rules \cite{Faustino}.
For example in the 
$SU(3)\times SU(2)\times U(1)_Y\times SO(10)_{hidden}$ model
of ref. \cite{Casas2}, whereas a large number of
renormalizable couplings are present, generating Fayet--Iliopoulos
scale masses $M_{FI}$ for the extra matter, only a small number 
of non-renormalizable couplings are allowed by gauge invariance.
Moreover, at the end of the day, the latter are forbidden by the
selection rules. 

Taking into account the above comments, we will adopt
in this paper the following point of view:
all the extra matter in the $Z_3$ orbifold models to be analysed
is massless or very heavy (with masses of the order of $M_{FI}$).
In the former situation they must acquire masses
through the electroweak symmetry breaking, as we will discuss below. 
In any case, 
since e.g. no new quarks have been observed
in colliders, their masses must be basically heavier than 200 GeV.
Again, in order to simplify the analysis, we will consider 
that the masses of these extra particles are of the order of $M_S$.

\section{Analysis of the RGEs for gauge couplings}

Let us turn now to the details of the calculation.
The one-loop runnings of the 
gauge couplings with energy $Q$ are
\bea
\frac{1}{\alpha_i(Q)}=\frac{1}{\alpha_i (M_Z)} 
+ \frac{b_i^{NS}}{2\pi}\ln\frac{M_{S}}{M_Z}
+ \frac{b_i^{S}}{2\pi}\ln\frac{M_{FI}}{M_S}
+ \frac{b_i^{FI}}{2\pi}\ln\frac{Q}{M_{FI}}
\ ,
\label{running}
\eea
where $\alpha_i=g_i^2/4\pi$ with $i=2,3,Y$, and
$b_i$'s are the coefficients of the
$\beta$-functions. In particular, using the matter content
of the standard model (with one Higgs doublet), the 
non-supersymmetric $\beta$-functions are given by
$b_3^{NS}=7$, $b_2^{NS}=19/6$ and $b_Y^{NS}=-C^2\times 41/6$.
As we will discuss in detail below, 
the normalization constant, $C$, of the $U(1)_Y$ hypercharge generator
is not fixed as in the case of grand unified theories (e.g. for
$SU(5)$, $C^2=3/5$), so for the moment we consider it as a free parameter.
The supersymmetric $\beta$-functions between the supersymmetric 
scale $M_S$ and the Fayet--Iliopoulos scale $M_{FI}$, 
considering three supersymmetric
generations of standard particles, two Higgs doublets and an
arbitrary number of extra particles are
\bea
b_3^S &=& 3-\frac{1}{2}n_3\ , 
\label{susybeta3}
\\
b_2^S &=& -1-\frac{1}{2}n_2\ ,
\label{susybeta2}
\\
b_Y^S &=& -C^2\times (11+q)\ , 
\label{susybeta1}
\eea
where 
\bea
q=\sum_{i=1}^{n_1} Y_i^2+2\sum_{j=1}^{n_2} Y_j^2+3\sum_{k=1}^{n_3}
Y_k^2\ ,
\label{qus}
\eea
and
$n_1$, $n_2$, $n_3$ is the number of extra $SU(3)\times SU(2)$ singlets, 
$SU(2)$ doublets and
$SU(3)$ triplets, respectively, with masses close to $M_S$ and 
hypercharges $Y_l$.
Finally, 
the supersymmetric $\beta$-functions beyond $M_{FI}$, 
considering an
arbitrary number of extra singlets, $n_1^{FI}$,
doublets, $n_2^{FI}$, and triplets, $n_3^{FI}$,
all of them with masses close to $M_{FI}$, are
%
\bea
b_3^{FI} &=& b_3^S-\frac{1}{2}n_3^{FI}\ , 
\\
b_2^{FI} &=& b_2^S-\frac{1}{2}n_2^{FI}\ ,
\\
b_Y^{FI} &=& b_Y^S-C^2\times (q^{FI})\ , 
\label{susyFIbeta}
\eea
where $q^{FI}$ is given by eq. (\ref{qus}) with the substitution
$n_{1,2,3}\rightarrow n_{1,2,3}^{FI}$.

It is worth remarking here that moduli-dependent string
threshold corrections to eq. (\ref{running}) are absent in the
case of the $Z_3$ 
orbifold \cite{Kaplu}.
Moduli-independent ones
are generically present but, although model-dependent, they are 
small \cite{Kaplu,Mayr} and we neglect them in the computation.

\begin{figure}[t]
\begin{center}
\begin{tabular}{c}
\epsfig{file= 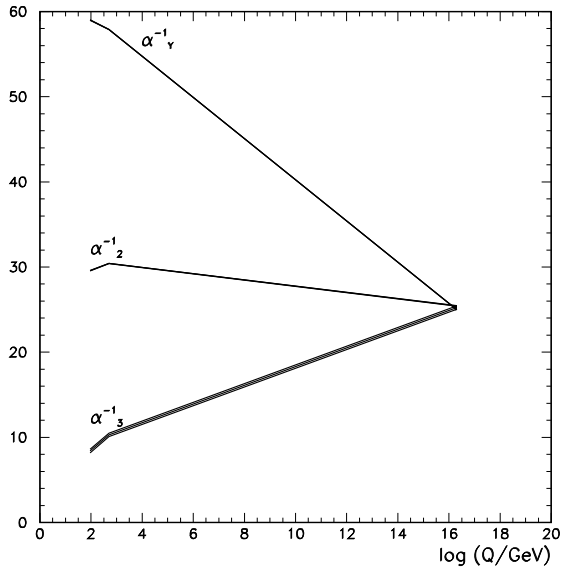, width=8.5cm}
\end{tabular}
\end{center} 
\caption{
Unification of the gauge couplings of the MSSM 
at $M_{GUT}\approx 2\cdot 10^{16}$ GeV, using $C^2={3/5}$
as the normalization factor for the hypercharge.
} 
\end{figure}

Now, using for example
the above equations with $n_3=n_2=q=n_3^{FI}=n_2^{FI}=q^{FI}=0$,
one is able to reproduce straightforwardly the
well-known prediction of the Minimal Supersymmetric Standard Model
(MSSM),
namely the three gauge couplings unify ({\it assuming} the
normalization constant of $SU(5)$)
at $M_{GUT}\approx 2\cdot 10^{16}$ GeV.
This is shown in Fig.~1, using
the experimental values \cite{pdg}
$M_Z=91.187$ GeV, $\alpha_3^{-1} (M_Z)=8.39\pm 0.2$, 
$\alpha_2^{-1} (M_Z)=29.567\pm 0.02$ and 
$\alpha_Y^{-1} (M_Z)={C^2}\times (98.333\pm 0.02)$, which are given
in the ${\overline{MS}}$ scheme. We neglect 
the conversion factors to the usual  ${\overline{DR}}$ supersymmetric
scheme since they are very small (see e.g. \cite{Langacker}). 
In 
addition we are neglecting higher-loop corrections to the
running (for an estimate of these small corrections see \cite{March}). 
The results shown in this and the other figures in this 
paper
are not going to be modified 
qualitatively by
these (small) effects. We have checked it explicitly (adding also 
moduli-independent string threshold corrections). 
For instance, even if two-loop effects spoiled
the unification, they could be counteracted by adjusting e.g. the scale
$M_S$.

As discussed in the Introduction, 
we are interested in the unification of the gauge couplings
at
$M_{GUT}\approx g_{GUT}\times 5.27\cdot 10^{17}$ GeV.
This is not a simple issue, and various approaches towards 
understanding it have been proposed in the literature \cite{Dienes}.
Some of these proposals consist of using string GUT models,
extra matter at
intermediate scales, heavy string threshold corrections, non-standard
hypercharge normalizations, etc.
In our case, we will try  
to obtain this value by using first the existence of extra matter at the scale
$M_S$. We will see that this is not sufficient and more
ideas must be involved.

\subsection{Predictions from the unification 
of 
$\alpha_3$ 
with 
$\alpha_2$}

Let us concentrate for the moment on
$\alpha_3$ and  $\alpha_2$, neglecting the 
scale $M_{FI}$. Recalling that three generations appear automatically 
for all the matter in $Z_3$ orbifold scenarios with two Wilson lines,
the most natural possibility
is to assume the presence of three light generations of supersymmetric
Higgses.
This implies that we have four extra Higgs doublets with respect to the
case of the MSSM and therefore we have to use
$n_2=4$ and $n_3=0$ in eqs. (\ref{susybeta2}) and (\ref{susybeta3}), 
respectively.
Unfortunately, this goes wrong.
Whereas $\alpha_3^{-1}$ remains unchanged,
the line of Fig.~1 for $\alpha_2^{-1}$ is pushed 
down (see eq. (\ref{running})). As a consequence, the two
couplings cross at a very low scale
($\approx 10^{12}$ GeV).
We could try to improve this situation by assuming the presence of
extra triplets in addition to the four extra doublets. 
Then the line for $\alpha_3^{-1}$ is
also pushed down and therefore the crossing might be obtained
for larger scales. However, even for the
minimum number of extra triplets that can be naturally obtained
in our scenario, $3\times \{(3,1)+(\bar 3, 1)\}$, i.e. $n_3=6$,
the ``unification'' scale turns out to be too large ($\approx 10^{21}$
GeV).
The next simplest possibility, $n_2=7$ and $n_3=6$, produces
the crossing at the scale $\approx 10^{15}$ GeV, again well below the required
value, as in our first attempt.
More extra triplets would imply at least $n_3=12$ and therefore  
$\alpha_3^{-1}$ becomes negative at the scale $\approx 10^{13}$ GeV.
Summarizing, using extra matter at $M_S$ we are not able to obtain the 
Heterotic String unification scale since
$\alpha_3$ never crosses $\alpha_2$ at
$M_{GUT}\approx g_{GUT}\times 5.27\cdot 10^{17}$ GeV.
Fortunately, this is not the end of the story. 
As we will show now,
the Fayet--Iliopoulos scale $M_{FI}$ is going to
play an important role in the analysis.

Using eq. (\ref{running}) with $i=3,2$, and imposing
$\alpha_3(Q=M_{GUT})=\alpha_2(Q=M_{GUT})$, one obtains the following value for
the unification scale, taking into account $M_{FI}$:
\bea
\ln\frac{M_{GUT}}{M_{FI}}=
\frac{
4\pi\left[\alpha_2^{-1} (M_Z)-\alpha_3^{-1} (M_Z)\right]
-\frac{23}{3}\ln\frac{M_S}{M_Z}
-\left[8+(n_2-n_3)\right]\ln\frac{M_{FI}}{M_S}
}
{
\left[8+(n_2-n_3)+(n_2^{FI}-n_3^{FI})\right]
}
\ .
\label{MI}
\eea
In order to determine whether or not the  
Heterotic String unification scale
can be obtained, we need to know the number of doublets and triplets
in our construction with masses of the order of the Fayet--Iliopoulos
scale $M_{FI}$ (in fact knowing the difference between
the number of doublets and the number of 
triplets, $n_2^{FI}-n_3^{FI}$, is enough). 
Let us explain this calculation in some detail.

In principle one can construct, within
the $Z_3$ orbifold with two Wilson lines, a number 
of the order of 50000 of three-generation 
models
with the
$SU(3)\times SU(2)\times U(1)^5$ gauge group 
associated to the first $E_8$
and the $(3,2)$ matter representation in the untwisted sector
(in case of the $(3,2)$ representation in the twisted sector, 
$SU(3)\times SU(3)\times U(1)^4$ is the smallest gauge group that
can be
obtained associated to the first $E_8$ \cite{Kimm}). 
However, a detailed analysis implied that most of them are equivalent.
In fact, at the end of the day, only 9 models are left \cite{Mondragon}.
The (observable) untwisted matter associated with the first $E_8$ is
uniquely determined and can be easily computed. There are only
two types.
Considering just the $SU(3)$ triplets and 
$SU(2)$ doublets, these are 
$3\times \{(3,2)\}$ and $3\times \{(3,2)+(\bar 3, 1)+(1,2)\}$ \cite{Mondragon}.
Thus there are three more doublets than triplets in both cases.
The determination of the
twisted matter of  
these models on a model-by-model basis is more involved, since 
one has to add several hidden-sector parts 
associated to the second $E_8$. This is necessary 
even for the twisted matter associated to the observable sector.
However, taking into account the result for the untwisted matter, 
duality-anomaly cancellation arguments show
that there must be nine more doublets than triplets in the 
observable twisted sector \cite{IL}.
Altogether one obtains that there are
twelve more doublets than triplets in all models.
Since the Standard Model (excluding the Higgs sector) contains
the same number of doublets as triplets,
$3\times \left\{(3,2)+2(\bar 3,1)+(1,2)\right\}$,
we obtain the relation $2+n_2+n_2^{FI}=n_3+n_3^{FI}+12$.
It is now straightforward to check, using eq. (\ref{MI}),
that only models with 
\bea
n_2=4\ ,\ n_3=6 
\ ,
\label{SM1}
\eea
%
and therefore 
$n_2^{FI}-n_3^{FI}=12$, may give rise to the Heterotic String
unification scale (the other possibilities for $n_2$, $n_3$, discussed
above do not even produce the crossing of $\alpha_3$ and $\alpha_2$). 
This is shown in Fig.~2. There we are
using $M_{FI}=2\cdot 10^{16}$ GeV as will be discussed below. We also
postpone for the moment the discussion about the coupling $\alpha_1$.

\begin{figure}[!t]
\begin{center}
\begin{tabular}{c}
\hspace{-1cm}
\epsfig{file= 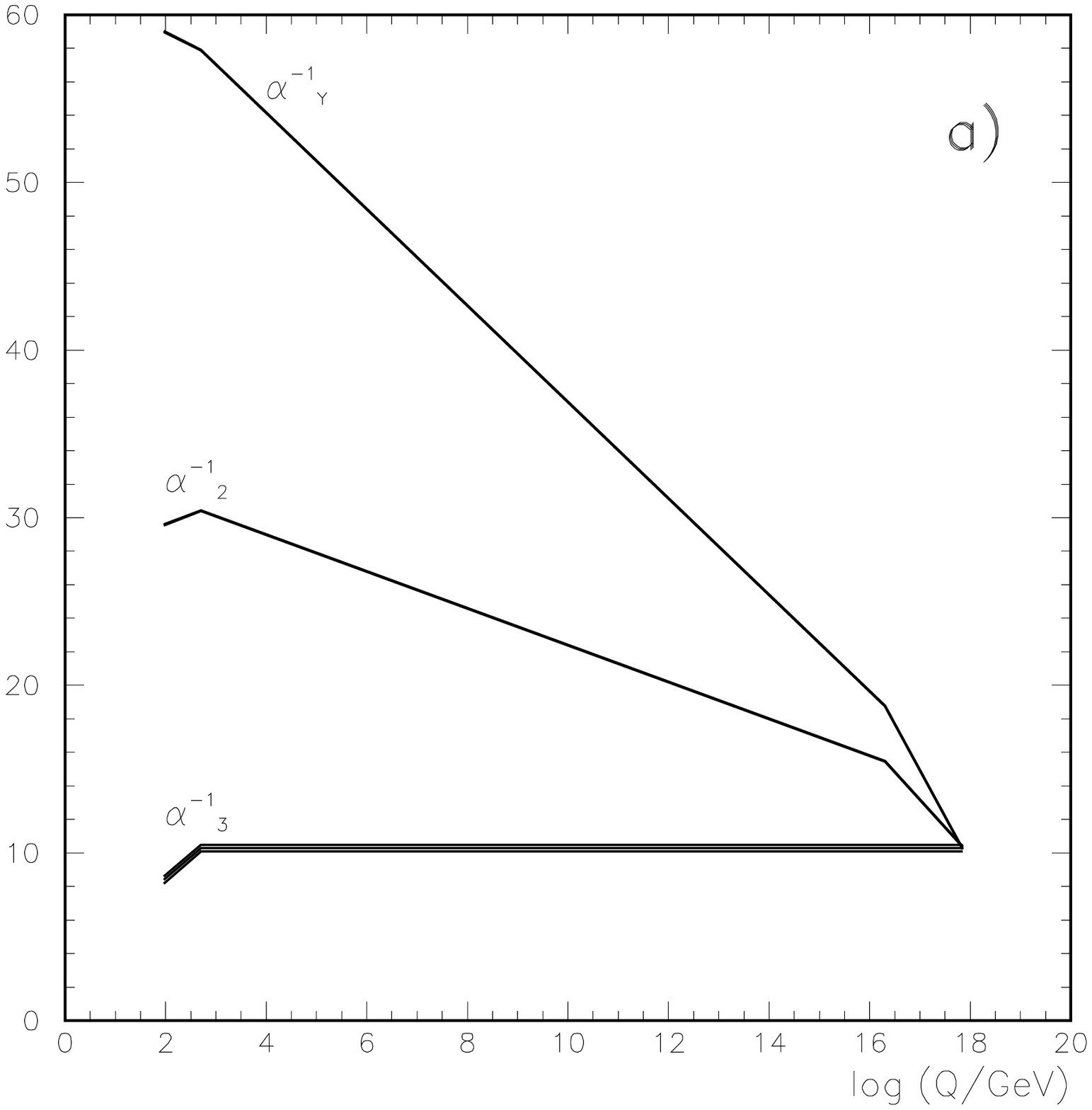, width=8.5cm}
\hspace{0.cm}\epsfig{file= 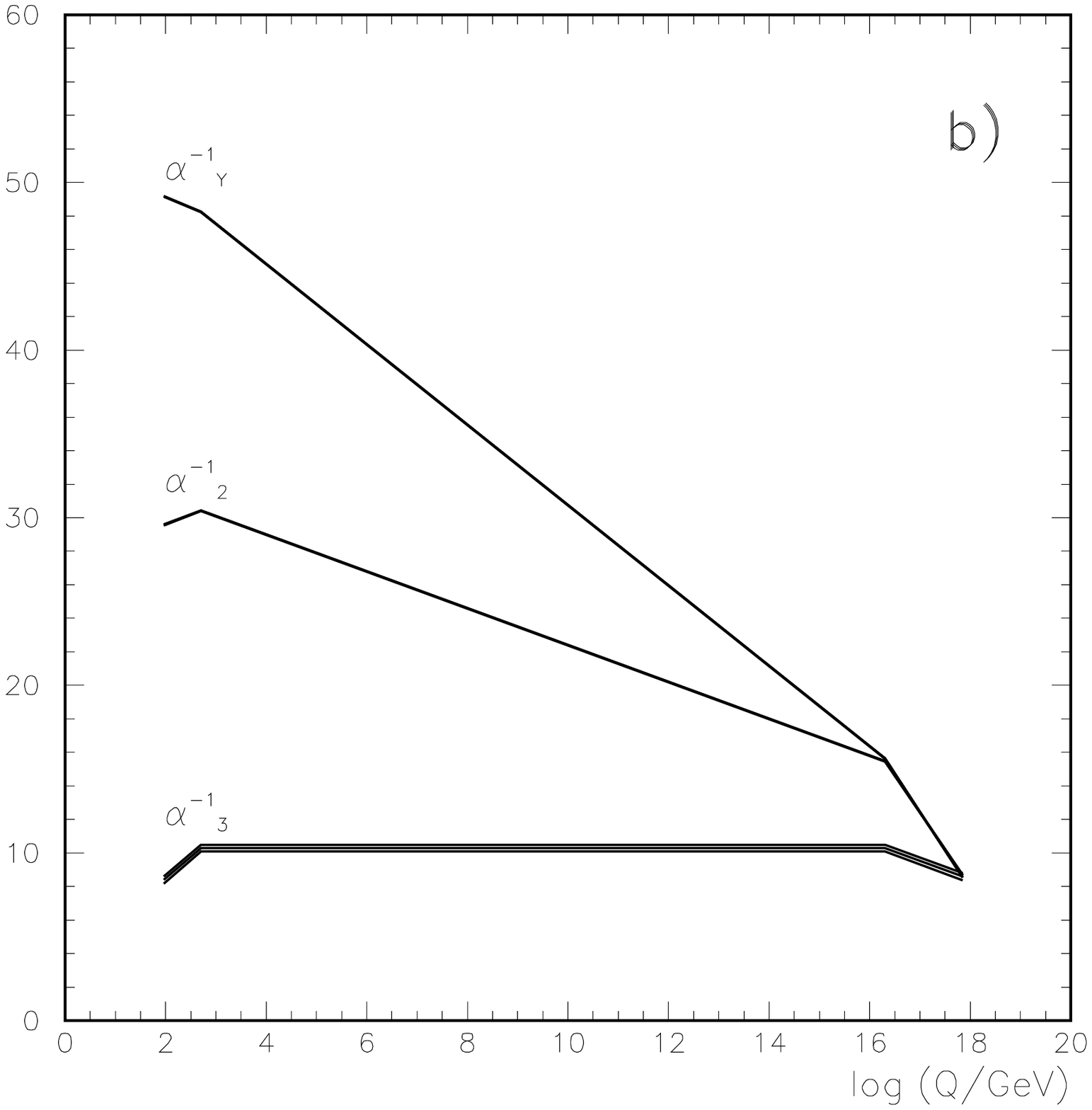, width=8.5cm}

\\
\hspace{-1cm}
\epsfig{file= 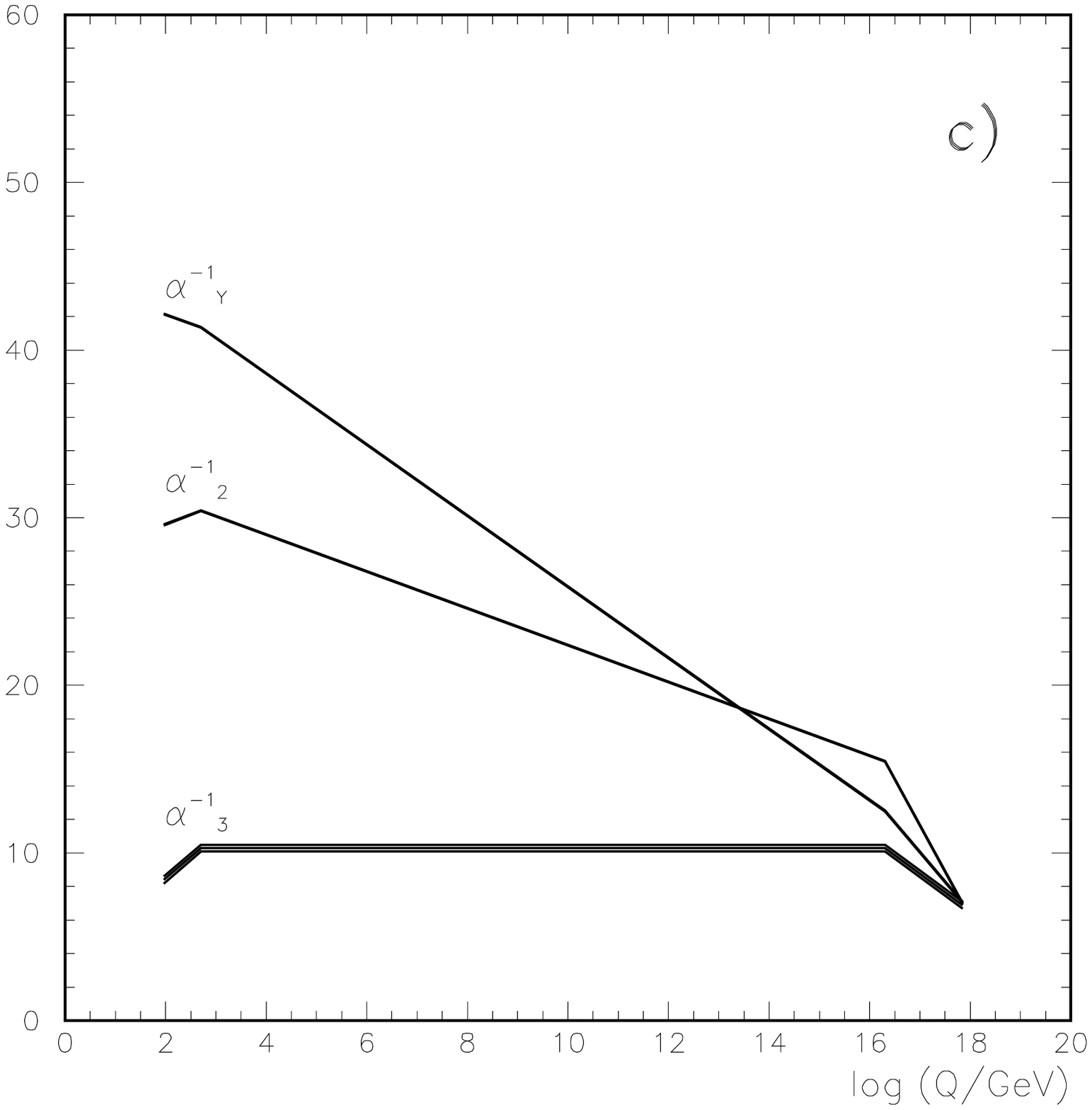, width=8.5cm}
\hspace{0.cm}\epsfig{file= 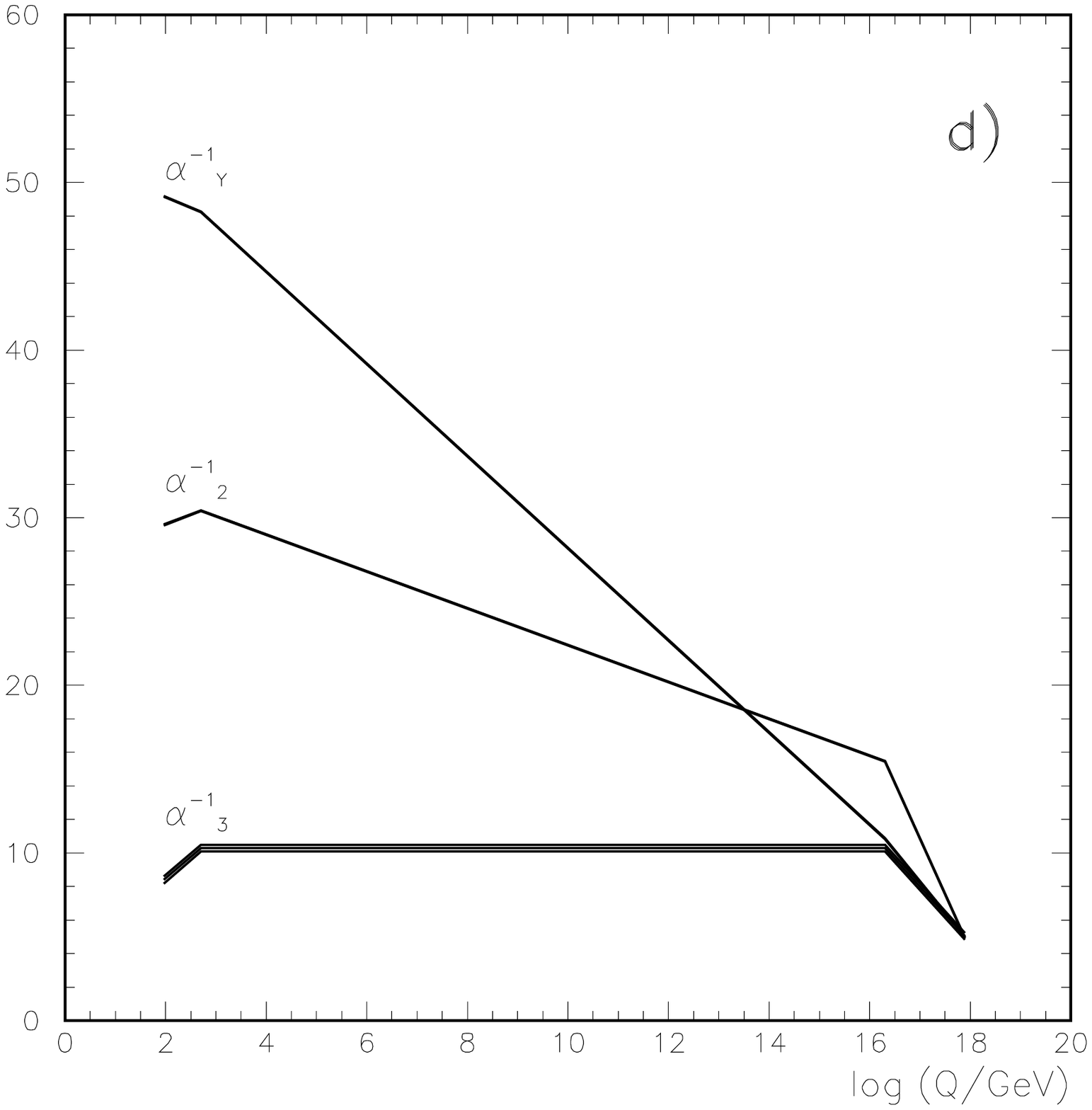, width=8.5cm}
\end{tabular}
\end{center} 
\caption{
Unification of the gauge couplings 
at $M_{GUT}\approx g_{GUT}\times 5.27\cdot 10^{17}$ GeV 
for the Heterotic String construction 
analysed in the text with three light generations of supersymmetric Higgses
and
vector-like colour triplets, $n_2=4$, $n_3=6$.
Cases a), b), c) and d) correspond to the four 
possible patterns of heavy matter in eq. (\ref{extra5}).
For each case, the line corresponding to $\alpha_1$
is just one example of the many possibilities discussed below eq. (\ref{C2}).}
\end{figure}

Note that at low energy we then have (excluding singlets)
\bea
3\times \left\{(3,2)+2(\bar 3,1)+(1,2)\right\}
+ 3\times \left\{(3,1)+(\bar 3,1)+ 2(1,2)\right\}
\ ,
\label{SM2}
\eea
i.e. the matter content of the Supersymmetric Standard Model with
three generations of Higgses and vector-like colour triplets.
In Section~4 we will discuss in some detail the phenomenology
associated to this scenario, where extra matter at low energy is present.


Let us remark that although in this scenario 
$M_{GUT}$ only depends on the difference between
the number of doublets and the number of 
triplets $n_2^{FI}-n_3^{FI}$, 
the 
value of the unified coupling constant $g_{GUT}$ indeed depends on
the precise number of matter multiplets.
For example, using eq.(\ref{running}) with $i=2$ we need to know
$n_2^{FI}$. 
Since $n_2^{FI}-n_3^{FI}=12$, the
following patterns of matter with masses of the order of $M_{FI}$
are allowed:
%
\bea
a)\ n_3^{FI}=0\ ,\,\ \ n_2^{FI}=12 & \rightarrow & 3\times \left\{4(1,2)\right\}
\ ,
\nonumber
\\
b)\ n_3^{FI}=6\ ,\,\ \ n_2^{FI}=18 & \rightarrow & 
3\times \left\{(3,1)+(\bar 3,1)+6(1,2)\right\}
\ ,
\nonumber
\\
c)\ n_3^{FI}=12\ ,\ n_2^{FI}=24 & \rightarrow &
 3\times \left\{2[(3,1)+(\bar 3,1)]+8(1,2)\right\}
\ ,
\nonumber 
\\
d)\ n_3^{FI}=18\ ,\ n_2^{FI}=30 & \rightarrow & 
3\times \left\{3[(3,1)+(\bar 3,1)]+10(1,2)\right\}
\label{extra5}
\ .
\eea
%
Patterns with 24 or more colour triplets 
(and the corresponding $SU(2)$ doublets) can be discarded.
The reason is the following. 
Recently, the analysis of ref. \cite{Mondragon} discussed above 
was extended in ref. \cite{Giedt}. 
There, the number of inequivalent models associated to the
first $E_8$
was reduced further and only 6 were left. 
In addition, when 
the second $E_8$
was added in the analysis, only 192 different models were found.
They have only five possible hidden-sector gauge groups,
$SO(10)\times U(1)^3$, 
$SU(5)\times SU(2)\times U(1)^3$,
$SU(4)\times SU(2)^2\times U(1)^3$,
$SU(3)\times SU(2)^2\times U(1)^4$,
$SU(2)^2\times U(1)^6$. Then,
the matter content of the 175 models 
associated to the first four gauge groups
was analysed in detail \cite{Giedt2}. It is straightforward to obtain
from that classification that for 162 of those models
only the
above four patterns are present. 
For the rest, either they have no anomalous $U(1)$ associated
(7 of them), and therefore the Fayet--Iliopoulos mechanism does not
work, or they have no extra triplets at all, $n_3=n_3^{FI}=0$ (6 of them).

For a given Fayet-Iliopoulos scale, $M_{FI}$, 
each one of the four patterns in eq. (\ref{extra5})
will give rise to a different value
for $g_{GUT}$. Adjusting $M_{FI}$ appropriately,
we can always get $M_{GUT}\approx g_{GUT}\times 5.27\cdot 10^{17}$ GeV.
In particular this is so for 
$M_{FI}\approx 2\times 10^{16}$ GeV as shown in Fig.~2. 
It is remarkable that this number is within the allowed
range 
for the Fayet--Iliopoulos breaking scale discussed below eq. (\ref{FI}).
We also see in Fig.~2 that patterns a), b), c) and d) have
$g\approx 1.1$, $g\approx 1.2$, $g\approx 1.3$ and
$g\approx 1.5$, respectively,
and therefore $M_{GUT}\approx 5.8\cdot 10^{17}$ GeV,
$M_{GUT}\approx  6.3\cdot 10^{17}$ GeV,
$M_{GUT}\approx 6.8\cdot 10^{17}$ GeV and
$M_{GUT}\approx 7.9\cdot 10^{17}$ GeV.

\subsection{Analysis of $\alpha_1$}

Of course, we cannot claim to have obtained the 
Heterotic String unification scale until we have shown
that the coupling $\alpha_1$ joins the other two couplings
at $M_{GUT}$.
The analysis becomes more involved now because we need to know
not only the hypercharges of the extra doublets and triplets
of our scenario but also the ones of the extra singlets.
In particular the latter are present in large numbers in these models
$\sim 3\times$(25--60).
Thus 
the analysis has to be carried out on a model-by-model basis.
In addition, as already mentioned above,
the normalization constant $C$ of the
$U(1)$ hypercharge generator is not fixed as in the case of
grand unified theories. In string constructions, the correct hypercharge for 
the physical particles is obtained as a combination of
$U(1)$'s \cite{Katehou}, $Y=\sum c_i U_i$, and therefore the
normalization factor is given by
$C=(\sum c_i^2)^{-1/2}$ \cite{rges}.
This means that the combination for the hypercharge depends on the
model
and, even for a specific model, there exist many acceptable 
combinations \cite{Katehou}.

The only model-independent information we have concerning the value of
$C$ is the upper bound found in ref. \cite{rges} for the case of 
$Z_3$ orbifold compactifications, namely $C\leq 1$. On the other hand,
the normalization factors of
the 175 models mentioned above were analysed in ref. \cite{Giedt2}
and it seems that only $C\lsim \sqrt{3/5}$ may be obtained.

Then, what we will try to do is to study whether or not it is
plausible, in these models, to obtain the correct value for
$\alpha_1$ at $M_{GUT}$.
Let us analyse first the value of $q$ in eq. (\ref{qus}). We know that 
the four extra doublets contribute with $q=2$, since they
are Higgses with hypercharges $\pm 1/2$.
On the other hand, we cannot know, without a model-by-model analysis,
the hypercharges associated to the six extra triplets.
The same argument applies to the extra singlets, which could also
be present at $M_S$. However, concerning the latter,
we have already argued above that, at least some of them, 
and in some explicit example all of them \cite{Font}, will
have vanishing hypercharge.
For the computation of $q^{FI}$ the situation is more uncertain,
since we do not even know the hypercharges associated to
the extra doublets.
So the only thing we can say about $q$ and $q^{FI}$ is that
they 
must fulfil the lower bounds $q > 2$, $q^{FI} > 0$.
Of course these bounds are very conservative. For example,
if we assume that all extra singlets left at the scale $M_S$ 
have $Y=0$, some of the other singlets with masses close to the
scale $M_{FI}$ will have non-vanishing values, and therefore 
will contribute to $q^{FI}$.


Using eq. (\ref{running}) with $i=Y$ we can compute,
for given values of $q$ and $q^{FI}$, the appropriate
value of $C^2$ in order to unify the coupling $\alpha_1$ with the
others at the Heterotic String unification scale $M_{GUT}$:
\bea
C^2=\frac{2\pi\times \alpha^{-1} (M_{GUT})}
{2\pi\times 98.333-
\frac{41}{6}\ln\frac{M_S}{M_Z}
-(11+q)\ln\frac{M_{FI}}{M_S}
-(11+q+q^{FI})\ln\frac{M_{GUT}}{M_{FI}}
}
\ .
\label{C2}
\eea
The results are shown in Fig.~3, where three possible values for $q$ are
considered, $q=2.08, 2.5, 4$, corresponding to six 
extra colour triplets (in addition to the four extra Higgs doublets)
at $M_S$, say $D$ and $\overline D$, 
with hypercharges $Y=\pm 1/15, \pm 1/6, \mp 1/3$, respectively.
These hypercharges for triplets appear in the three $Z_3$ orbifold models 
with two Wilson lines studied in detail 
in the literature. In particular, extra triplets with
$Y=\mp 1/3$ appear in the model of ref. \cite{Casas2},
with $Y=\pm 1/6$ (and also with $Y=\mp 1/3$) in the model 
of ref. \cite{Font}, and with $Y=\pm 1/15$ in \cite{Giedt2}. 
Triplets with hypercharge $\pm 2/3$ also appear in the models of
refs. \cite{Casas2,Font}.
However, their contribution would imply a large value for $q$,
in particular $q=10$, and therefore no solution for
positive $C^2$ can be found, even with $q^{FI}=0$. 
Thus it would be better to give, through the Fayet--Iliopoulos mechanism, 
heavy masses for those triplets (contributing only to $q^{FI}$).

\begin{figure}[t!]
\begin{center}
\begin{tabular}{c}
\hspace{-1cm}
\epsfig{file= 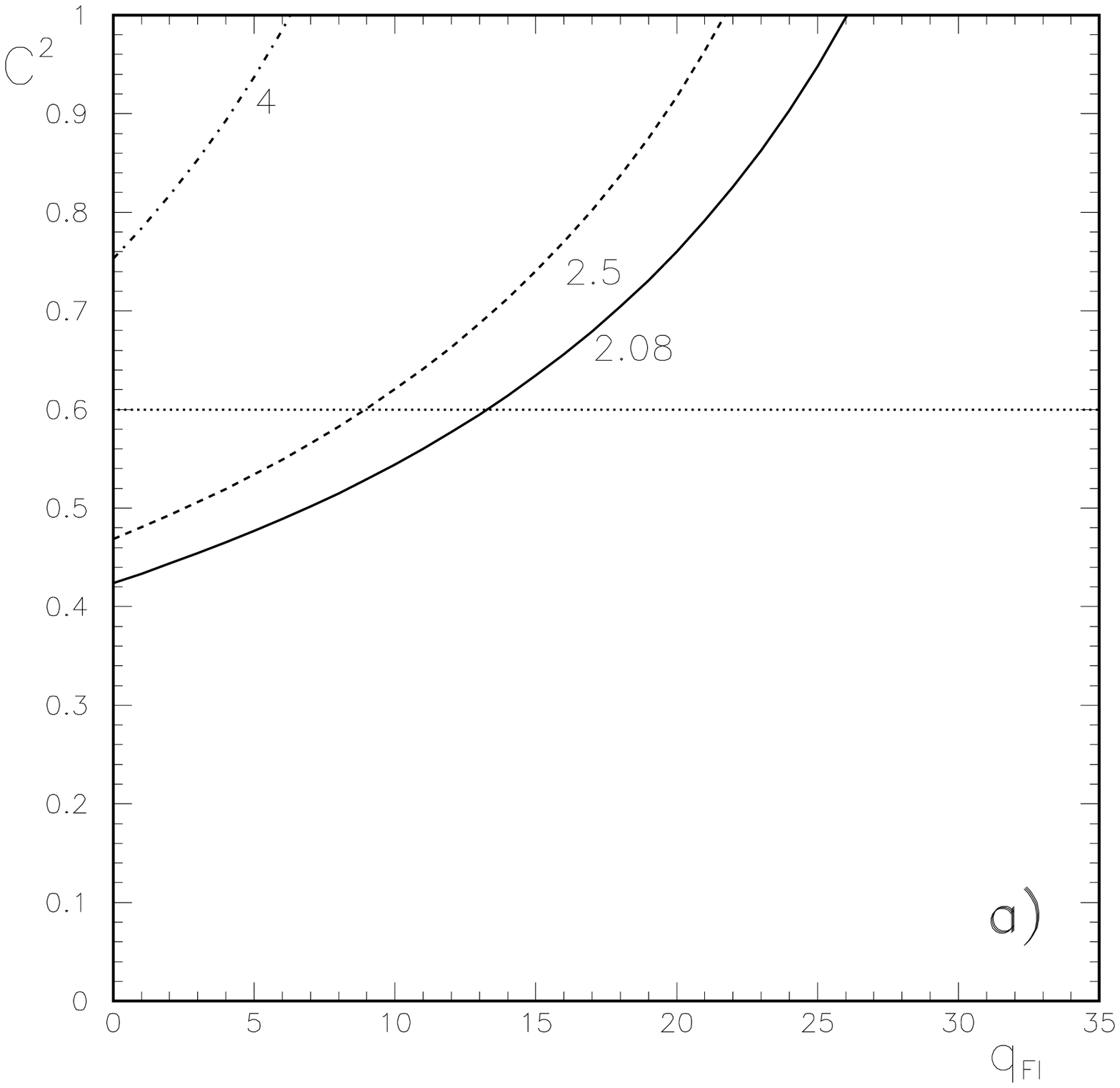, width=8.5cm}
\hspace{0.cm}\epsfig{file= 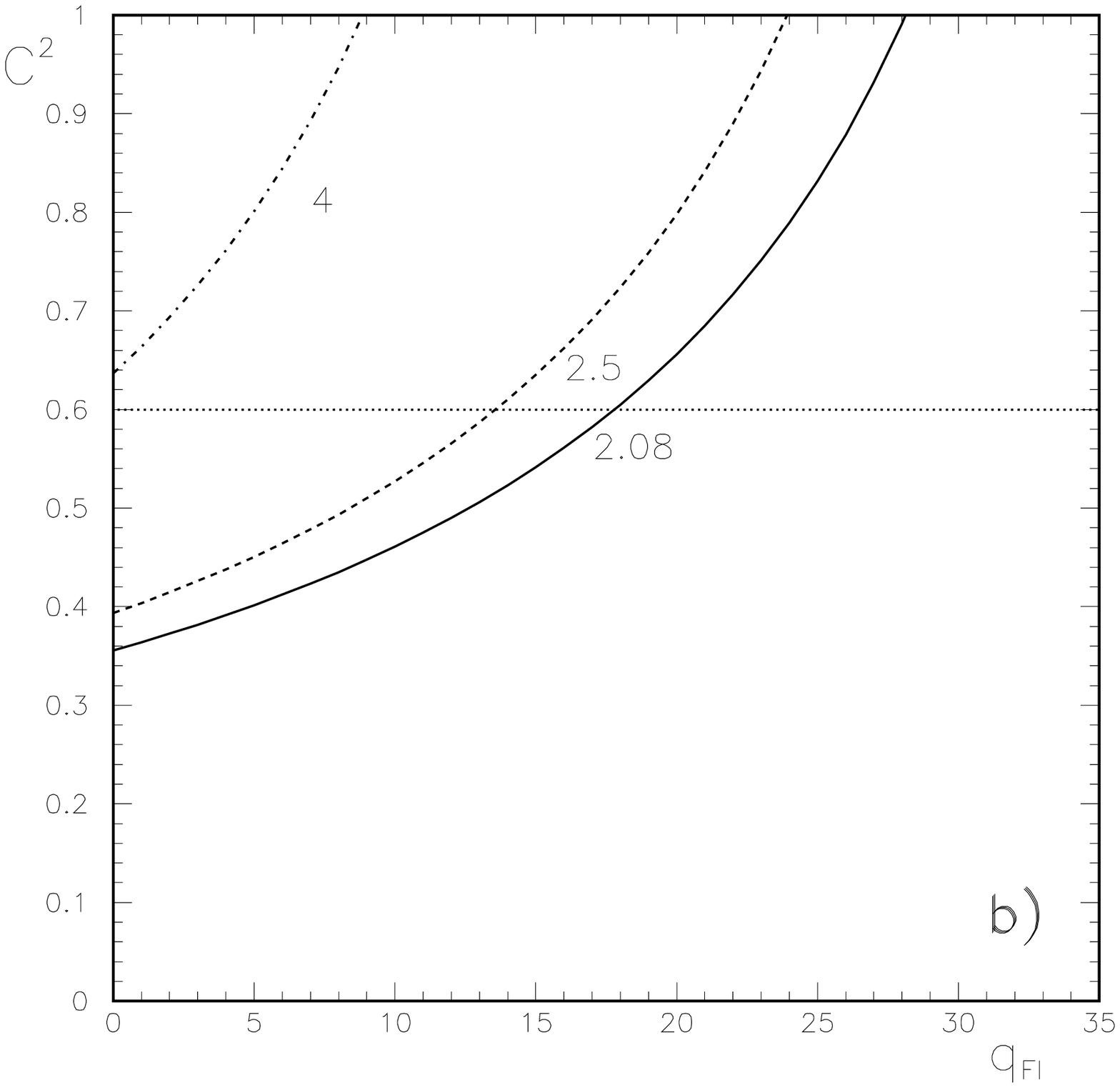, width=8.5cm}

\\
\hspace{-1cm}
\epsfig{file= 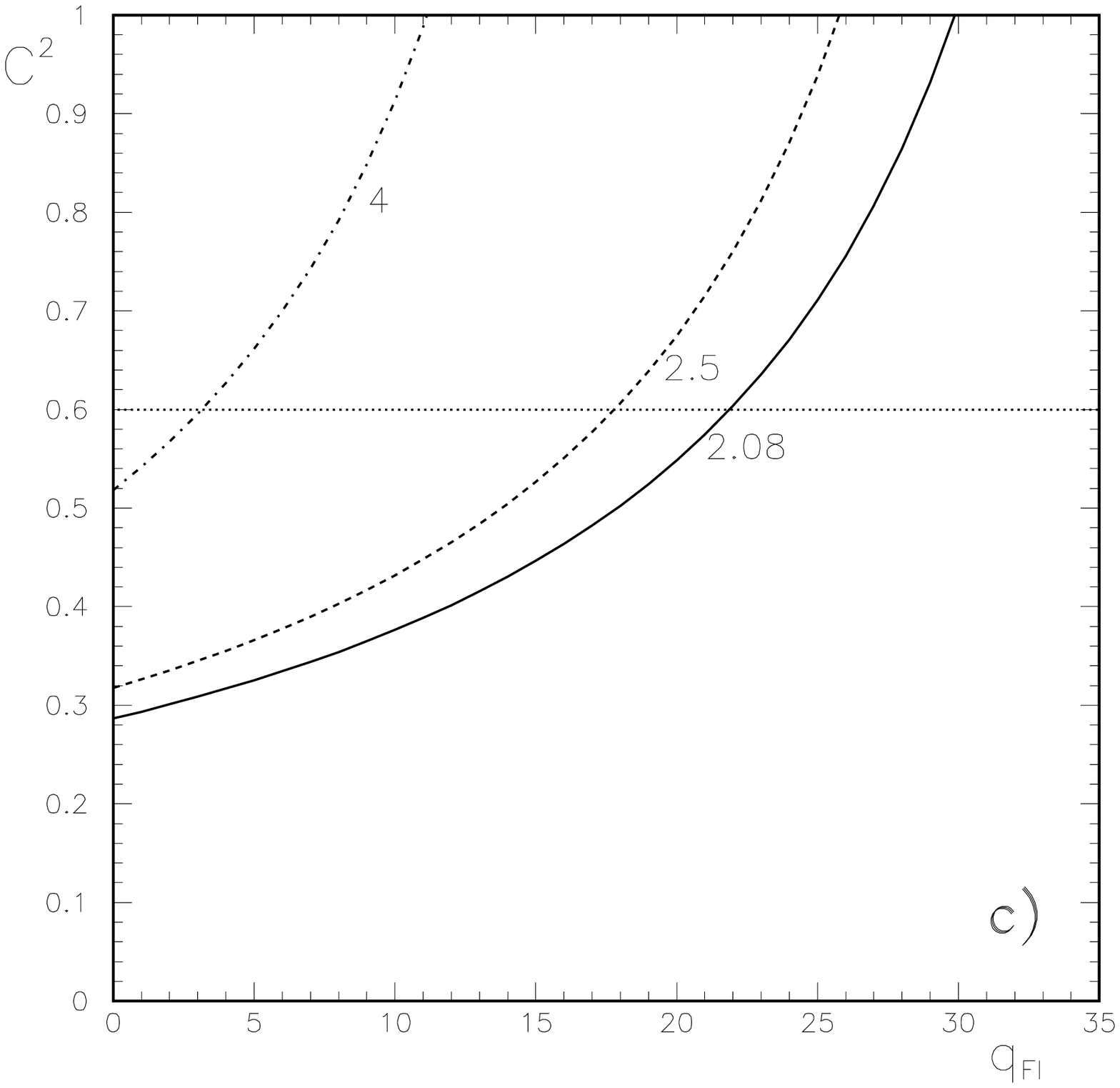, width=8.5cm}
\hspace{0.cm}\epsfig{file= 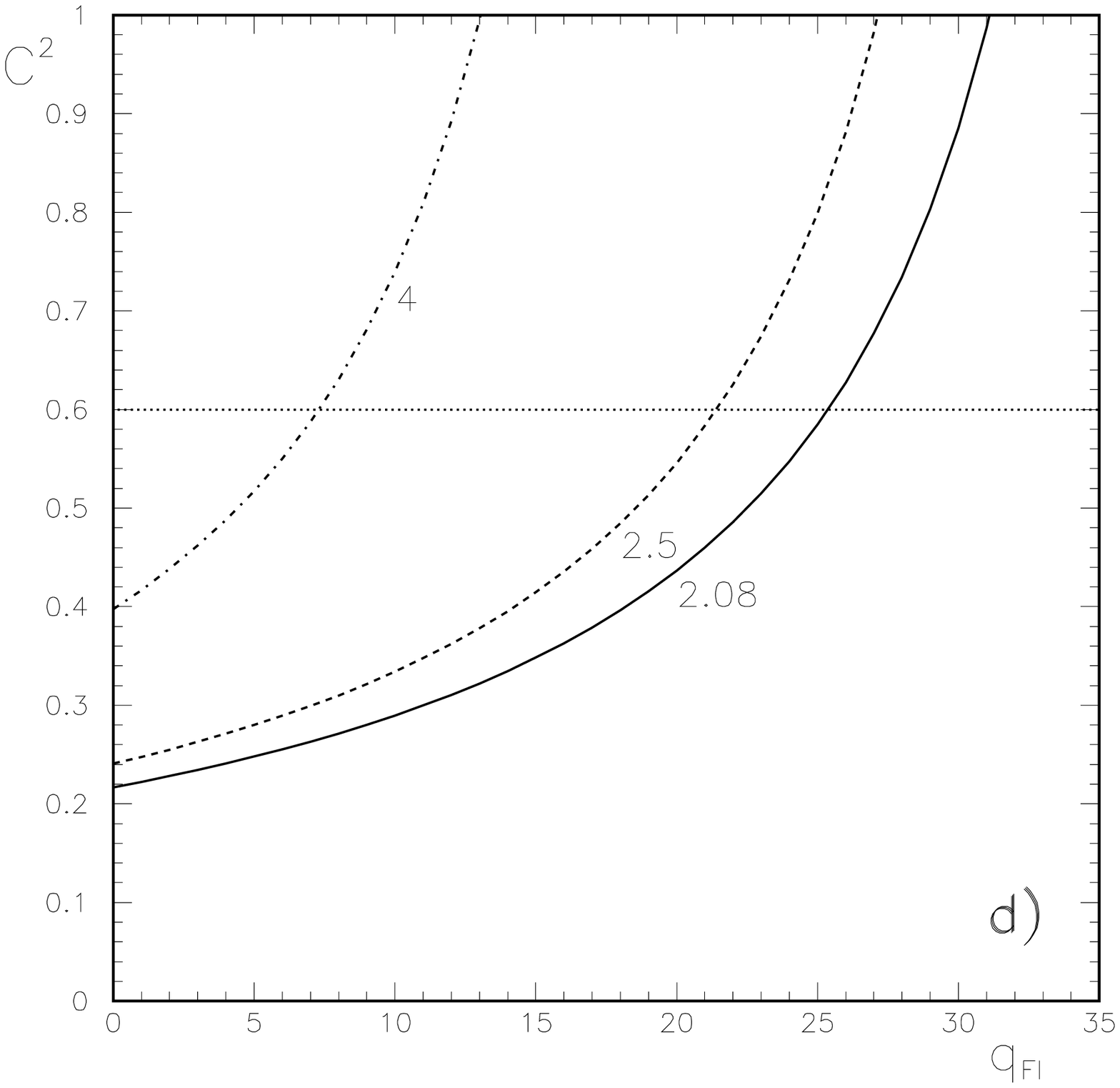, width=8.5cm}
\end{tabular}
\end{center} 
\caption{Normalization factor of the hypercharge $C^2$ versus
$q^{FI}$ in order to obtain the Heterotic String unification scale,
for three different values of $q$.
As discussed in the text we are using $q=2.08, 2.5, 4$.
The straight line
indicates the usual $SU(5)$ normalization,
$C^2=3/5$.
The results for the four possible patterns of heavy extra matter 
in eq. (\ref{extra5}), a), b), c) and d), are shown.
} 
\end{figure}

From Fig.~3a we obtain that pattern a) implies the following lower bound
for the normalization constant: $C^2\gsim 3/7$. For instance,
for $q=2.08$ and $q^{FI}=12$ we need to have $C^2\approx 3/5$ in order 
to unify $\alpha_1$ with the other couplings. This example is the
one shown in Fig.~2a. The lower bound associated to pattern b)
is  $C^2\gsim 3/9$. For the above example we would now need 
$C^2\approx 3/6$, and this is shown in Fig.~2b. To pattern c)
is associated the bound $C^2\gsim 3/11$. For instance,
for $q=2.5$ and $q^{FI}=9$ we need $C^2\approx 3/7$. This is shown
in Fig.~2c. Finally, for pattern d) the bound is  $C^2\gsim 3/15$.
In Fig.~2d we show the case  $q=4$, $q^{FI}=5$, where 
$C^2\approx 3/6$ must be used.

The simplicity of the constraints that we have found is extremely
useful in order to perform a systematic analysis of the phenomenological
viability of all possible vacua. It is very likely that most of them
can easily be discarded. Good examples are the above 
mentioned models \cite{Casas2,Font,Giedt2}.
In ref. \cite{Katehou} two possible $U(1)$ combinations for the
hypercharge were obtained for the model introduced in ref. \cite{Kim}.
One of them, with $C^2=3/17$ \cite{rges}, was studied in detail \cite{Casas2}.
Assuming condition (\ref{SM1}) after the 
Fayet--Iliopoulos breaking, the model corresponds to pattern
d) in eq. (\ref{extra5}), with 
$q=4$, $q^{FI}=99$. 
We can check in Fig.~3d
that there is no solution because of the large value of $q^{FI}$.
In other words, $C^2$ becomes negative using eq. (\ref{C2}),
and therefore unification of the coupling $\alpha_1$ with the others
is not possible.
In ref. \cite{Font} the other combination with $C^2=3/11$ was studied.
Its analysis yields $q=2.5$, $q^{FI}=64.5$.
Again, because of the large value of $q^{FI}$, unification is not possible.
Finally, let us consider the model studied in ref. \cite{Giedt2}
with $C^2=15/37\approx 0.4$,
which, assuming condition (\ref{SM1}), would correspond
to pattern a). Its analysis yields $q=2.08$, $q^{FI}=21.2$.
Now, as can be seen from Fig.~3a, 
a solution for $C^2$ producing unification is possible,
namely $C^2\approx 0.76$. Unfortunately, the latter does not coincide
with the normalization of the model written above.
Furthermore, as discussed above, $C^2$ is unlikely to
be much bigger than 0.6.

It is important to remark again that the combination for the
hypercharge is not unique in these orbifold constructions \cite{Katehou}.
There are a lot of choices and some of them could satisfy the
constraints that we have found. Anyway, even if this is not
the case for these models, there are other 177 models that
have not yet been analysed in detail. 

\section{Phenomenology of this scenario}

The main characteristic of the scenario studied in previous sections,
is the presence at low energy of extra matter. In particular, we have
obtained that three generations of Higgses and vector-like colour
triplets are necessary.

Since more Higgs particles than in the MSSM are present,
there will be of course a much richer phenomenology \cite{cero}--\cite{tres}.
Note for instance that the presence of six Higgs doublets
implies the existence of sixteen physical Higgs bosons,
eleven of them are neutral and five charged.  
On the other hand, it is well known that dangerous
flavour-changing neutral currents (FCNCs) may appear when
fermions of a given charge receive their mass through couplings
with several Higgs doublets \cite{Paschos}.
This is because the transformations diagonalizing the fermion
mass matrices do not, in principle, diagonalize the Yukawa interactions.
This situation might be present here since we have three generations
of supersymmetric Higgses.
In general, the most stringent limit on flavour-changing processes comes
from the small value of the $K_L-K_S$ mass difference \cite{KK}.
There are two approaches in order to solve this potential problem.
In one of them one assumes that the extra Higgses are
sufficiently  massive making 
$\Delta S=2$ neutral currents 
small enough not to contradict
the experimental data \cite{Georgi,KK,Cheng}. In this case
the actual lower bound 
on Higgs masses depends on the particular texture
choosen for the Yukawa matrices, but can be as 
low as 120--200 GeV \cite{fcnc}.
In the other approach the 
Yukawa couplings have some symmetries eliminating
FCNCs completely \cite{Paschos}. 
The simplest example is when the couplings between the extra Higgses 
and quarks of a 
given charge are forbidden. If the three Yukawa-coupling matrices
are present, still one can avoid FCNCs 
if the matrices are proportional. In this case one can always choose a basis
in which only one generation of Higgses couples to quarks.
It would be very interesting to analyse which approach arises
naturally in these orbifold models thanks to
the stringy selection rules \cite{who?}.

Concerning the three generations of vector-like colour triplets, 
$D$ and $\overline D$, 
they should acquire masses above the experimental limit ${\cal O}$(200
GeV).
This is possible, in principle,
through 
couplings with some of the extra singlets with $Y=0$, say $N_i$, 
which are usually 
left at low energies, even after the Fayet--Iliopoulos breaking.
For example, in the model of ref. \cite{Casas2}, there are
13 of these singlets.
Thus couplings $N_iD\overline D$ might be present.
From the electroweak symmetry breaking, the fields $N_i$
a VEV might
develop. Note in this sense that the Giudice--Masiero 
mechanism to generate a $\mu$ term through the
K\"ahler potential is not available in prime orbifolds as $Z_3$ \cite{Gava}.
Thus an interesting possibility to generate it, 
given the large number of singlets present in orbifold models, 
is to consider couplings of the type $N_jH_uH_d$.
It is also worth noticing that some of these singlets might not have
the necessary couplings to develop
VEVs and then might be candidates for right-handed neutrinos.

Before concluding, a few comments about the hypercharges of the
extra colour triplets are necessary. Of the three models that we
have used as examples in the previous section,
two of them, the ones with 
$C^2=15/37$ and $C^2=3/11$, 
have triplets with non-standard fractional electric charge,
$\pm 1/15$ and $\pm 1/6$ respectively.
The existence of this kind of matter is a generic property
of the massless spectrum of supersymmetric models \cite{Wen1,Wen2}.
This means that they have necessarily 
colour-neutral fractionally charged states, since
the triplets bind with the ordinary quarks.
For example, the model with triplets with electric charge
$\pm 1/6$ will have mesons and baryons with charges 
$\pm 1/2$ and $\pm 3/2$.
On the other hand,
the model with $C^2=3/17$ 
has `standard' extra triplets, i.e. with electric charges
$\mp 1/3$ and $\pm 2/3$; these will therefore give rise
to colour-neutral integrally charged states.
For example, a $d$-like quark $D$ forms
states of the type $u\overline D$, $uuD$, etc.
These results are consistent with general arguments \cite{Schelle}:
level-one string models can be modular-invariant and free of colour-neutral
fractionally charged states if and only if
\bea
\frac{7}{12}+\frac{C^{-2}}{4}=0\ (mod\ 1)
\ .
\label{charge}
\eea
It is trivial to see that only the model with $C^2=3/17$
fulfils this condition.

Let us now briefly discuss if our assumption of light
extra colour triplets to solve the unification
problem is consistent with the above results.
In principle, the existence of stable charged states creates
conflicts with cosmological bounds. 
For example, thermal production 
of these particles would overclose the Universe unless
their masses are below a few TeV \cite{Coriano,Perl}.
In models with
`non-standard' extra triplets, the lightest 
colour-neutral fractionally charged state,
due to electric charge conservation,
will be stable. However, since its mass must be 
of the order of the supersymmetric scale $M_S$
to have unification, the above mentioned conflict is not present in our 
case.
On the other hand, as pointed out in ref. \cite{Wen2},
the estimation of its relic abundance
contradicts limits on the existence of fractional charge in matter
(less than $10^{-20}$ per nucleon \cite{Perl}).
Thus, avoiding such fractionally charged states is necessary.
A possible mechanism to carry it out is inflation. Inflation would
dilute these particles, saving these unification models.
The reheating temperature $T_{RH}$ should be low enough not to
produce them again.
A recent calculation implies that $T_{RH}$ must be smaller than
$10^{-3}$ times the mass of the particle \cite{Riotto}, i.e. 
$T_{RH}\lsim 1$ GeV
in our case.
This is possible in principle, since the only constraint on this
temperature is to be larger than 1 MeV not to spoil
the successful nucleosynthesis predictions. 
Another possibility studied in the literature \cite{hidden}
to solve the problem is that
the extra triplets transform also under a
non-Abelian group in the hidden sector. Then, they
may be confined into integrally charged states.
However, in the orbifold models studied here this possibility
is not available, since the extra triplets are always singlets
under the non-Abelian hidden groups.

Let us concentrate now on the `standard' extra triplets.
In addition to the possible mass terms, $ND\overline D$,
discussed above, these triplets could have FCNC couplings with ordinary 
quarks, e.g. $H_d Q \overline D$. Then decays
of the $D$'s through couplings with Higgses and quarks are possible,
and therefore the colour-neutral integrally charged 
states will not be stable.
Other decay channels are present in this case.
It is well known that FCNC couplings may appear if all fermions
with the same charge and helicity do not have the same 
$SU(2)$ quantum numbers, because of their mixing through mass 
matrices \cite{Paschos,Paschos2}.
This is true even for one family of Higgses.
As a consequence, 
the $D$'s can also decay 
through charged and neutral currents. 
Thus models where this type of triplets
are light ($\sim M_S$) may unify the couplings at $M_{GUT}$ without
any cosmological problem,
and will only be observed in collider experiments.
Analyses of their production modes and decays will be similar to
those of extra fermions in $E_6$ theories \cite{e6}.
Of course, there are further dangerous operators involving the $D$'s,
like $QL\overline D$, $\bar u \bar d \overline D$, etc., 
which must be forbidden by $U(1)$ gauge invariance or
stringy selection rules.
Recall, as discussed below eq. (\ref{C2}), that light
triplets with hypercharge $\pm 2/3$ are not good 
to obtain the correct value for $C^2$, so only light $D$'s with
hypercharge $\mp 1/3$ will be helpful.
Note that the above model with $C^2=3/11$ may also belong to this class
since, in addition to triplets with electric charge $\pm 1/6$, it also
has triplets with charge $\mp 1/3$, and these could be the light ones.

On the other hand, the colour-neutral integrally charged states
are not automatically unstable in all models. The appropriate Yukawa
couplings
may be forbidden by $U(1)$ gauge invariance or stringy selection
rules, and therefore these states would be stable.

\section{Final comments and outlook}

We have attacked the problem of the unification of gauge couplings in
Heterotic String constructions. Assuming that the Standard Model arises
from the $Z_3$ orbifold, we have obtained that 
$\alpha_3$ and $\alpha_2$ cross at the right scale, in a natural way,
when a certain type of extra matter is present. In this sense 
three families of supersymmetric Higgses and
vector-like colour triplets might be observed in forthcoming experiments.
The unification with 
$\alpha_1$ is obtained if the model has the
appropriate normalization factor of the hypercharge.
Our solution implies that the apparent unification of the MSSM using
$SU(5)$ normalization is just an accident, without any physical relevance.

Let us recall that
although we have been working with explicit orbifold examples, our arguments
are quite general and can be used for other 
schemes where
the Standard Model gauge group with three generations of particles
is obtained,
since extra matter and
anomalous $U(1)$'s are generically present in compactifications of the
Heterotic String.
Even models with gauge groups larger than that of the Standard Model
might be analyzed following the lines of this paper,
after their breaking, using e.g. the Fayet-Iliopoulos term, to 
$SU(3)\times SU(2)\times U(1)_Y\times G_{hidden}$.

Once we have guessed the right phenomenological properties that,
in our opinion, the candidates to the Superstring Standard Model must
have,
the next natural step consists in searching explicitly the right
model \cite{who2?}. 
The main difficulty resides in how to obtain the
observed structure of
fermion masses and mixing angles. It is true that
one can find interesting results in the literature.
In particular, orbifold spaces have a beautiful mechanism to generate
a mass hierarchy at the renormalizable level.
Namely, Yukawa couplings can be explicitly computed 
and they get suppression factors, which depend on the
distance between the fixed points to which the relevant fields are
attached \cite{Faustino,nuevos}. 
These distances can be varied by giving different
VEVs to the $T$-moduli associated to the size and shape of the orbifold.
Of course, as usual in String Theory, it is extremely difficult
to implement this mechanism in a particular model.
However, we argue again that if the Standard Model arises from this type of
constructions, 
there must exist one model where this can be done. 
Perhaps we may turn the above difficulty into a virtue, since the weird
structure of Yukawas might be used as a hint to find the
model.

If, at the end of the day, a model with the characteristics described
above is found, this would be
a great success. Of course it is not the end of the story, since
in order to compute the explicit values of Yukawas we need to know
the VEVs of the $T$-moduli. Unfortunately, these are related
to the breaking of supersymmetry, and this is one of the 
biggest problems in String Theory. As a matter of fact we should
also be able to compute the value of $g_{GUT}$ 
obtained through phenomenological arguments in Section~3. But, again,
this is given by the VEV of another modulus field ($S$),
and therefore it is also connected with the mechanism of supersymmetry
breaking.
It is true that there are candidates for this task, such as gaugino
condensation in a hidden sector, and that we have hidden gauge groups
that could condensate. However, again, implementing this mechanism
in a particular model is not easy.

Finally, the soft terms should be computed in order to connect the
Supersymmetric Standard Model with the low-energy world. They should
fulfil all kinds of phenomenological requirements. For instance,
avoiding dangerous FCNCs is one of them. $Z_3$  
orbifold models with two Wilson lines automatically fulfil 
it, 
since the three generations of a given type of particle are associated
to the same sector.

In conclusion, the task to be carried out in order to find the Superstring
Standard Model is very hard, cumbersome, and in some sense, tedious,
but indispensable if one wants to show that String Theory is
the fundamental theory of particle physics.

\bigskip

\noindent {\bf Acknowledgements}

\noindent We gratefully acknowledge J.A. Casas
for illuminating discussions and for his collaboration during the
early stages of this work. We also thanks D.G. Cerde\~no
for his invaluable help.

This work has been supported 
in part by the CICYT, under contract FPA2000-0980, and
the European Union, under contract 
HPRN-CT-2000-00148.

\newpage

\end{document}